\documentclass[10pt,prl,amsmath,amssymb,aps,showpacs,twocolumn,unsortedaddress]{revtex4-1}
\def\be#1\ee{\begin{equation}#1\end{equation}}
\def\ba#1\ea{\begin{align}#1\end{align}}

\newcommand{\lr}[1]{\left( #1\right)}

\newcommand{\dd}{\mathrm{d}}
\renewcommand{\vec}[1]{\mathbf{#1}}

\newcommand{\eqqref}[1]{\text{Eq.}\,\eqref{#1}}
\pdfoutput=1

\usepackage{hyperref}
\usepackage{graphicx}
\usepackage[normalem]{ulem}
\usepackage[T1]{fontenc}
\usepackage{dcolumn}
\usepackage{bm}
\usepackage[caption=false]{subfig}
\usepackage{slashed}
\usepackage{pdfpages}
\makeatletter
\AtBeginDocument{\let\LS@rot\@undefined}
\makeatother
\begin{document}

\title{\boldmath Nonrelativistic fermions with holographic interactions and the unitary Fermi gas}

\author{N.W.M. Plantz}
\author{H.T.C. Stoof}
\affiliation{Institute for Theoretical Physics and Center for Extreme Matter and Emergent Phenomena, \\Utrecht University, Princetonplein 5, 3584 CC Utrecht, The Netherlands}

\email{n.w.m.plantz@uu.nl}
\email{h.t.c.stoof@uu.nl}

\date{\today}

\begin{abstract}
We present an alternative way of computing nonrelativistic single-particle spectra from holography. To this end, we introduce a mass gap in a holographic Dirac semimetal and subsequently study the nonrelativistic limit of the resulting spectral functions. We use this method to compute the momentum distributions and the equation of state of our nonrelativistic fermions, of which the latter can be used to extract all thermodynamic properties of the system. We find that our results are universal and reproduce many experimentally and theoretically known features of an ultracold Fermi gas at unitarity.
\end{abstract}

\pacs{11.25.Tq, 03.75.Ss}

\maketitle
\flushbottom
\textit{Introduction.}--- Our understanding of ultracold Fermi gases has significantly progressed over the past decade, due to the fact that the $s$-wave scattering length, which is the relevant measure for the strength of the interactions in these systems, can be conveniently engineered by tuning a magnetic field near a so-called Feshbach resonance \cite{Tiesinga,Tiesinga2}. This allows for an accurate experimental analysis of ultracold gases in both the weakly and strongly coupled regime \cite{Jin,Zwierlein,Jin2,Salomon,Chin,Hulet}. A particularly interesting situation occurs exactly at resonance, where the external magnetic field is such that the scattering length diverges. At this point collisions between the atoms are unitarity limited and the system becomes almost scale invariant, in the sense that the only length scale at zero temperature is the average interatomic distance that is set by the atomic density and diverges at zero density. Consequently, the thermodynamic properties of the Fermi gas become universal at unitarity \cite{Ho}.
	
Being strongly coupled, close to scale invariant and experimentally accessible, these ultracold gases at unitarity present a benchmark problem for the application of the holographic AdS/CFT correspondence, which aims to describe a (possibly deformed) conformal field theory (CFT) as a boundary property of a dual theory in a curved spacetime with one more spatial dimension \cite{Zaanen2}. This correspondence was discovered within string theory \cite{Maldacena1999} and for condensed-matter physics has especially had some successes in the application to emergent relativistic systems such as graphene \cite{Sachdev1,Sachdev2} and Weyl or Dirac semimetals \cite{HartnollLec,JacobsUndoped,Landsteiner1,Landsteiner2,Jacobs3,Landsteiner3,Liu1,Liu2}. A common way to deal with nonrelativistic systems in holography is to use instead of an anti-de Sitter (AdS) spacetime background a so-called Lifshitz background \cite{Son1,McGreevy1,Kachru1,Taylor1,Sybesma} as a gravitational dual with a dynamical exponent $z=2$. However, the fermionic spectra obtained in this way are generally particle-hole symmetric and without a mass gap. Hence for the description of an ultracold gas of massive atoms, a different approach is needed. The purpose of this Letter is to provide this alternative approach to nonrelativistic holography, which allows us to compute nonrelativistic single-particle spectra that can in principle be compared with experiments. Our method uses as its starting point results for the dynamics of Dirac fermions from holography \cite{Plantz2}, from which we can also obtain single-particle spectra with a mass gap by introducing a mass deformation in the CFT. The introduction of the mass gap allows us to consider the nonrelativistic limit of such spectra, where this mass scale, which contains the speed of light $c$, is large compared to all the other energy scales in the problem. Our most important finding below is that we obtain a data collapse for the spectral functions in the limit $c \rightarrow \infty$, i.e., the spectral functions are universal after an appropriate scaling with the chemical potential.

\begin{figure}[!t]
	\includegraphics[trim = 0cm 0.1cm 0cm 0.1cm, clip=true, scale=.55]{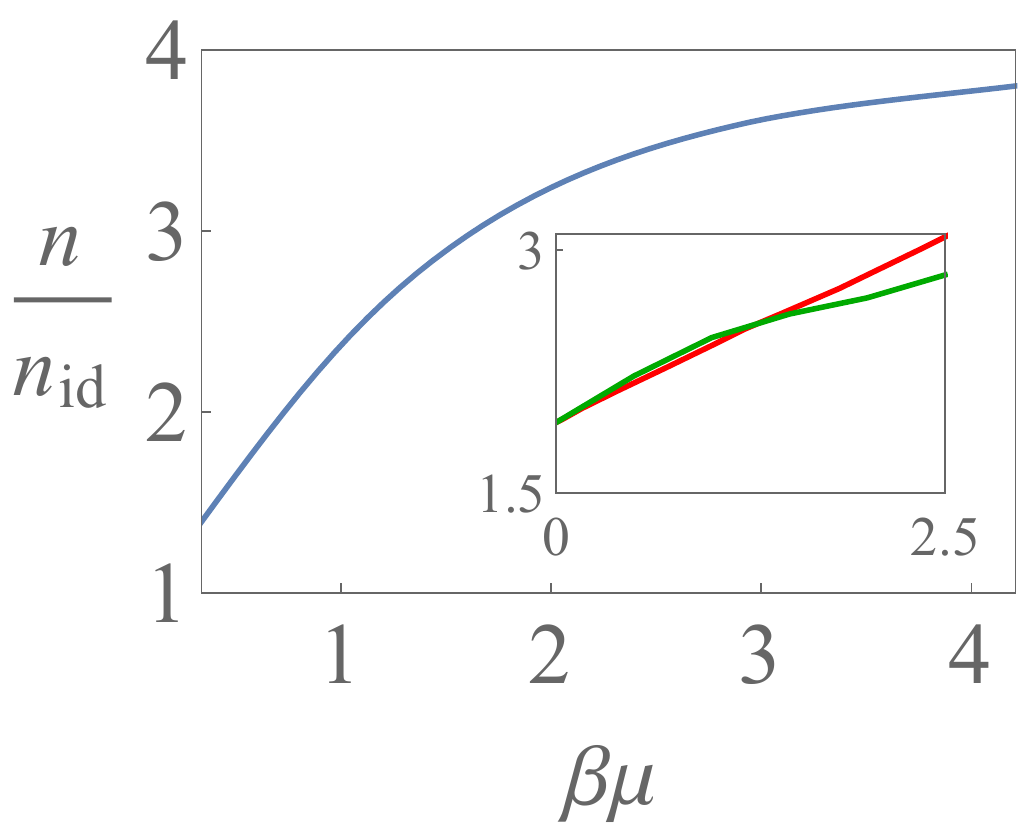}
\caption{(color online) The universal equation of state obtained for our nonrelativistic fermions with holographic interactions. The atomic density $n$ divided by the ideal Fermi gas density $n_{\text{id}}$ is shown as a function of the chemical potential times the inverse thermal energy $\beta\mu \equiv \mu/k_{\rm B} T$.  For comparison the inset shows the results of Refs.~\cite{Ku} (red) and \cite{Rammelmuller} (green). Note that our construction of the holographic interactions depends on four dimensionless model parameters that are introduced in the text below and that are not yet fine-tuned to the experiments as these contain effects of the inhomogeneity of the harmonically trapped gas and of the phase transition to the superfluid state. The specific values used are here the same as in Fig.\ 3.}\label{fig:density}
\end{figure}

An important advantage of our procedure is that it allows us to directly compare results obtained from a holographic model with experimental data. Therefore, we also extensively discuss the application of our method to ultracold Fermi gases at unitarity. In particular, we determine the equation of state from the single-particle spectra, i.e., the density as a function of the chemical potential and temperature, which is shown in Fig. \ref{fig:density} and from which all thermodynamic functions follow. The equation of state can be directly compared with results from experiments \cite{Schafer,Ku,Thomas,Salomon2} and from other theoretical models that are based on for example quantum Monte-Carlo methods \cite{Burovski1,Astrakharchik}, the Luttinger-Ward formalism \cite{Zwerger1}, functional renormalization-group methods \cite{Gubbels}, or more recently the complex Langevin model \cite{Rammelmuller}.

\textit{Holographic interactions.}--- To explain most clearly the physical content of our approach, we consider a relativistic Dirac fermion $\Psi$ with bare mass $M_0$ and chemical potential $\mu_0$ that is linearly coupled to a strongly interacting CFT through a fermionic operator $\mathcal{O}$. Referring to the supplemental material for our conventions in this section on the units (mostly $\hbar=c=1$) and on the Dirac theory in flat and curved spacetimes \cite{Note2}, the corresponding grand-canonical action is
\be
S=\int_k \left\{\Psi^\dagger\gamma^0\lr{-\slashed{K}-iM_0}\Psi+g\Psi^\dagger \mathcal{O}+g\mathcal{O}^\dagger\Psi\right\} + S_{\text{CFT}},
\ee
with $\int_k\equiv \int\dd^4k/(2\pi)^4$, $k_\mu=(-\omega,\vec{k})$, $\slashed{K}=\gamma^\mu K_\mu$, $\gamma^\mu$ the gamma matrices, $K_\mu=(-\omega-\mu_0,\vec{k})$, $g$ a coupling constant and $S_{\text{CFT}}$ the action of the deformed CFT containing $\mathcal{O}$. To make a connection with condensed-matter physics, we think of the CFT as being formed out of collective variables of the single fermion $\Psi$. From this perspective, the operator $\mathcal{O}$ is then a composite operator containing $\Psi$. The CFTs described by holographic models contain a large number of degrees of freedom $N$ \cite{Maldacena1999}, which implies that upon integrating out the CFT we can write the retarded Green's function for $\Psi$ as
\be
G_R^{-1}(\omega,\vec{k})=-\gamma^0\lr{\slashed{K}+iM_0}-\Sigma(\omega,\vec{k}),
\ee
with $\Sigma(\omega,\vec{k}) \equiv g^2G_{\mathcal{O}}(\omega,\vec{k})$ the self-energy matrix for $\Psi$ that due to the implicit large-$N$ limit only involves the two-point function $G_\mathcal{O}$ of $\mathcal{O}$. The latter can be directly obtained from the dictionary of the AdS/CFT correspondence.

\begin{figure}[!t]
\subfloat[\label{fig:profiles}]
	{\includegraphics[trim = 0cm 0cm 0cm 0cm, clip=true, scale=.475]{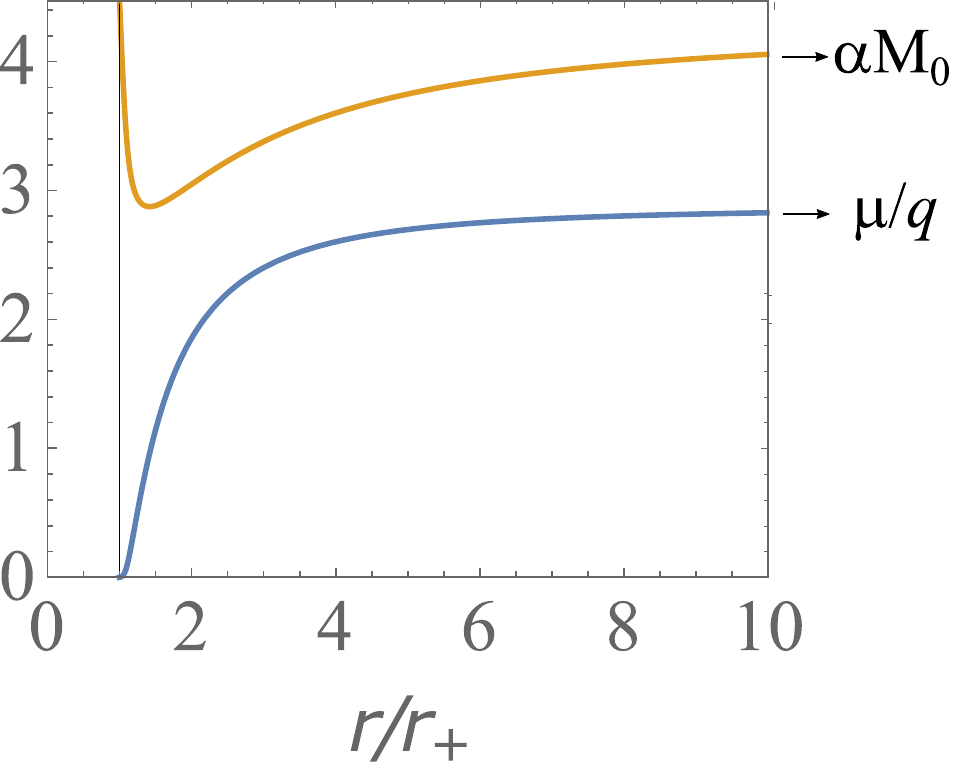}}
\subfloat[\label{fig:probe}]
	{\includegraphics[trim = 0cm 5.8cm 0cm 1.5cm, clip=true, scale=.086]{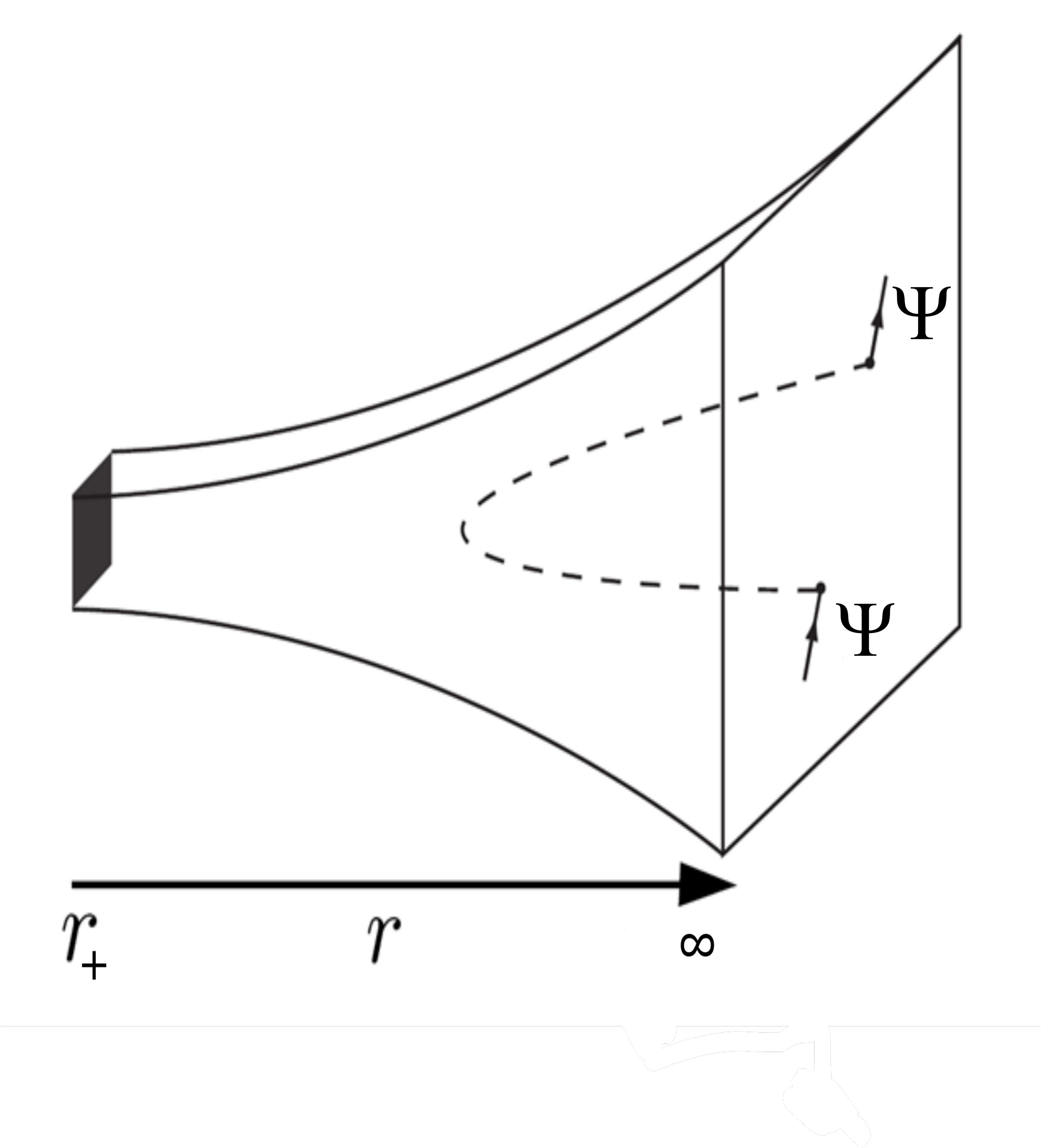}}
\caption{(color online) (a) A typical bulk solution of the gauge field $A_t$ (blue) and the scalar field $r\phi$ (yellow). The latter is multiplied by $r$ so that its value at the boundary at $r=\infty$ gives the mass $M_0$, up to the constant $\alpha=\sqrt[4]{3/\pi^2}$ that is derived in Ref. \cite{Plantz2}. From the value of $A_t$ at the boundary we can read off the chemical potential divided by the charge $q$. (b) The Witten diagram from which the self-energy for $\Psi$ follows. The dashed line gives the propagator $G_{\mathcal{O}}$, which follows from the propagation of the Dirac fermions in the curved bulk spacetime.}
\label{fig:background}
\end{figure}

Technically, we obtain the above Green's function $G_R$ from a holographic dynamical-source model \cite{JacobsUndoped}. The calculation of the Green's function is then a two-step process. The first step is to find the gravitational dual of the CFT, also known as the bulk background, which consists of a so-called asymptotically anti-de Sitter spacetime with an additional spatial coordinate $r$. Moreover, to have a nonzero temperature $T$ and chemical potential $\mu_0$ in the CFT, we need to have a black-hole horizon at $r=r_+$ and a $U(1)$ gauge field $A=A_t\dd t$ in the bulk \cite{HartnollLec}. Finally, consistent with our above interpretation of $\mathcal{O}$, we need to introduce a mass deformation in the CFT. This we achieve by adding also a real scalar field $\phi$ to the gravity theory \cite{Landsteiner1}. The gravitational background is then found by simultaneously solving the Einstein equations, the Maxwell equations and the Klein-Gordon equation. Numerically, this is achieved by integrating the coupled equations of motion for $A_t(r)$, $\phi(r)$ and the metric $g_{MN}(r)$, or equivalently the vielbeins $e^M_N(r)$ \cite{Note1}, from the horizon at $r_+$ to the boundary at $r=\infty$, where the CFT lives \cite{Note3}. Here we use capital Roman indices in the five-dimensional bulk spacetime, which, as opposed to the Greek indices, include the radial $r$-direction. From the boundary values of the solution we can then read off the chemical potential and the mass $M_0$, as illustrated in Fig. \ref{fig:profiles}, whereas the temperature is equal to the Hawking temperature that follows from the behavior of the metric at the horizon.

The second step is then to find the two-point function $G_\mathcal{O}$, that according to the holographic dictionary follows by having two Dirac spinors, which together contain the degrees of freedom of $\Psi$ and $\mathcal{O}$, propagate on the gravitational background found in the first step, as illustrated in Fig. \ref{fig:probe}. These spinors have bulk charge $q$ under the $U(1)$ gauge field and bulk masses $M$ and $-M$, respectively. Furthermore, they are coupled to the scalar field $\phi$ by a Yukawa coupling with strength $\lambda$, which is necessary to provide a coupling between the chiral components of the boundary spinor $\Psi$ \cite{Plantz2}. From the associated equations of motion for these bulk fermions, we can then derive a differential equation for the $4\times 4$ matrix $\Xi$, which is related to $G_\mathcal{O}$ by
$
G_{\mathcal{O}}(\omega,\vec{k})=-\lim_{r\rightarrow\infty}r^{2M}\gamma^0\Xi(r,\omega,\vec{k}).
$
This equation reads
\be \label{eq:xi}
-(e_r^r\partial_r+2M)\Xi+i\lr{i\slashed{K}+\lambda\phi}-i\Xi\lr{i\slashed{K}-\lambda\phi}
  \Xi=0,
\ee
where now $K_\mu=(-\omega-qA_t,\vec{k})$ and $\slashed{K}=\gamma^\nu e^\mu_\nu K_\mu$. It is supplemented with the initial condition $\Xi(r_+)=i\gamma^0$, corresponding to purely infalling conditions at the horizon. Having solved \eqqref{eq:xi}, we find the spectral function $\rho(\omega,\vec{k})=-\text{Im[Tr\,}G_R(\omega,\vec{k})]/\pi$ of $\Psi$ which depends on the ratios $k_{\rm B}T/M_0c^2$ and $\mu_0/M_0c^2$ obtained from the gravitational background, and additionally on the dimensionless parameters $q$, $M$, $\lambda$ and $g$ involved in our construction of the holographic interactions. We comment on the physical significance of these model parameters at the end of the paper.

\textit{Nonrelativistic limit.}---
\begin{figure}[!t]
\subfloat[\label{fig:nrlim}]
	{\includegraphics[trim = 0cm 0cm 0cm 0cm, clip=true, scale=.165]{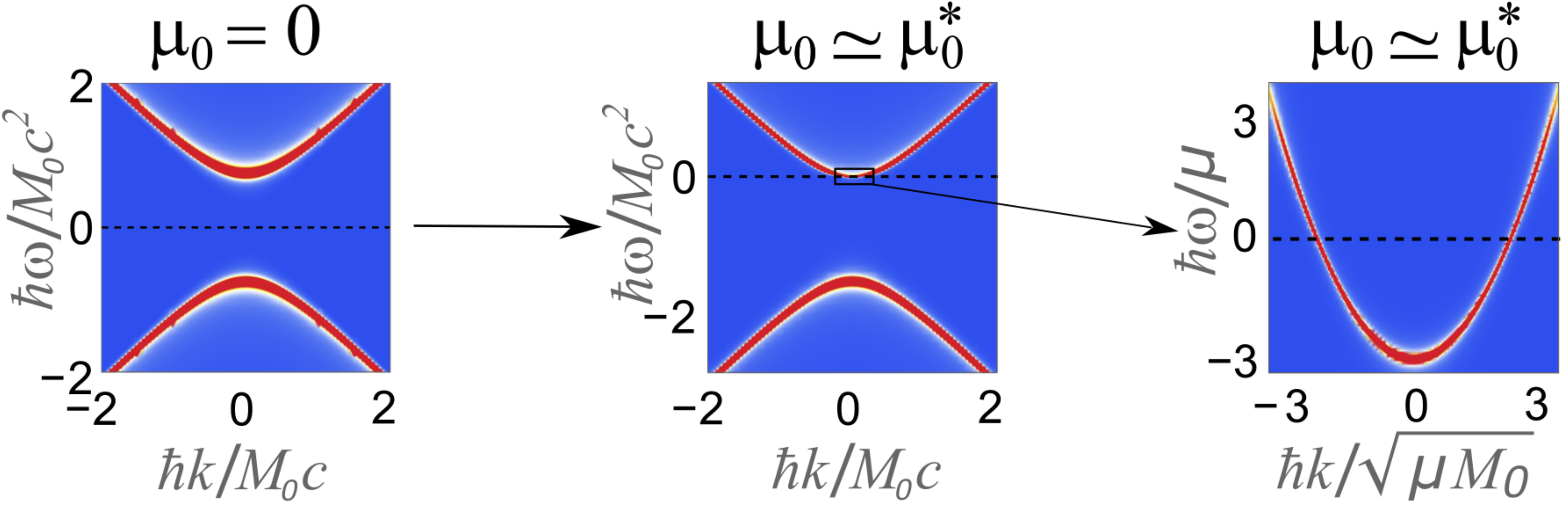}}

\subfloat[\label{fig:collapse}]
	{\includegraphics[trim = 0cm -0.64cm 0cm 0cm, clip=true, scale=.37]{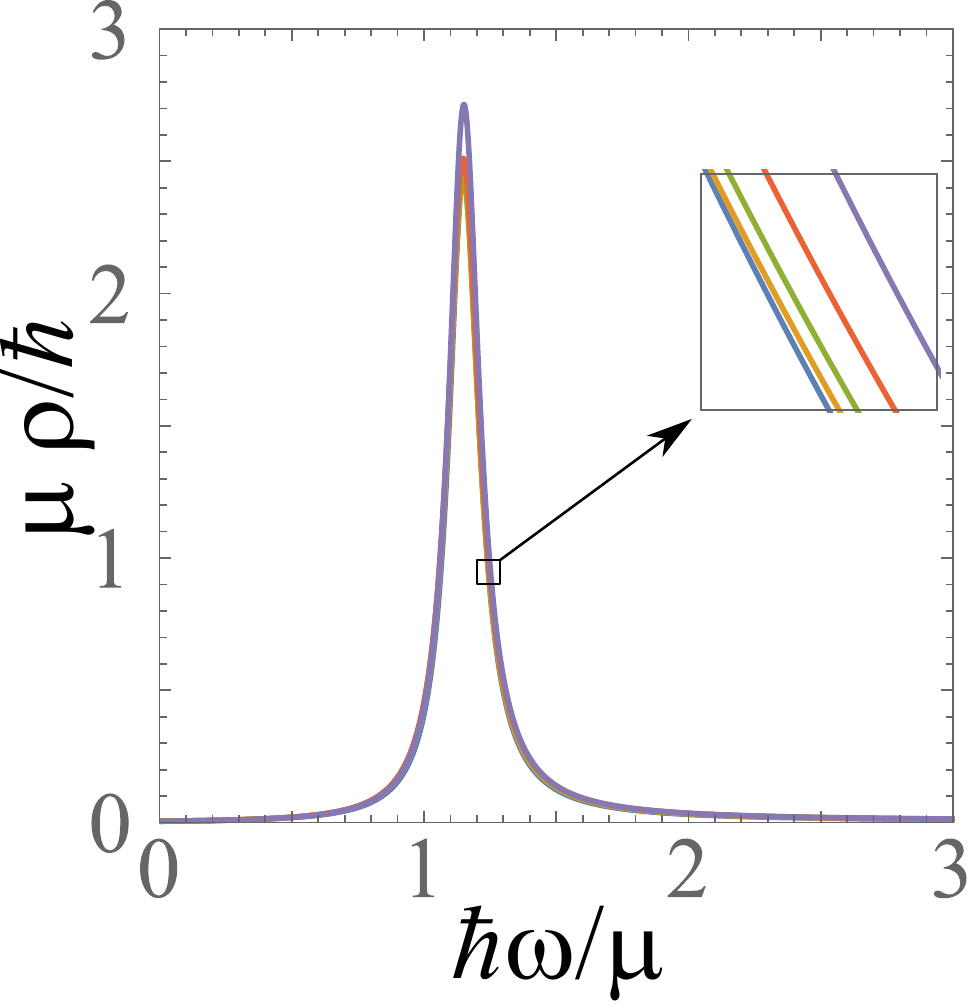}}
\subfloat[\label{fig:spectralfunctions}]
	{\includegraphics[trim = 0cm 0cm 0cm 0cm, clip=true, scale=.396]{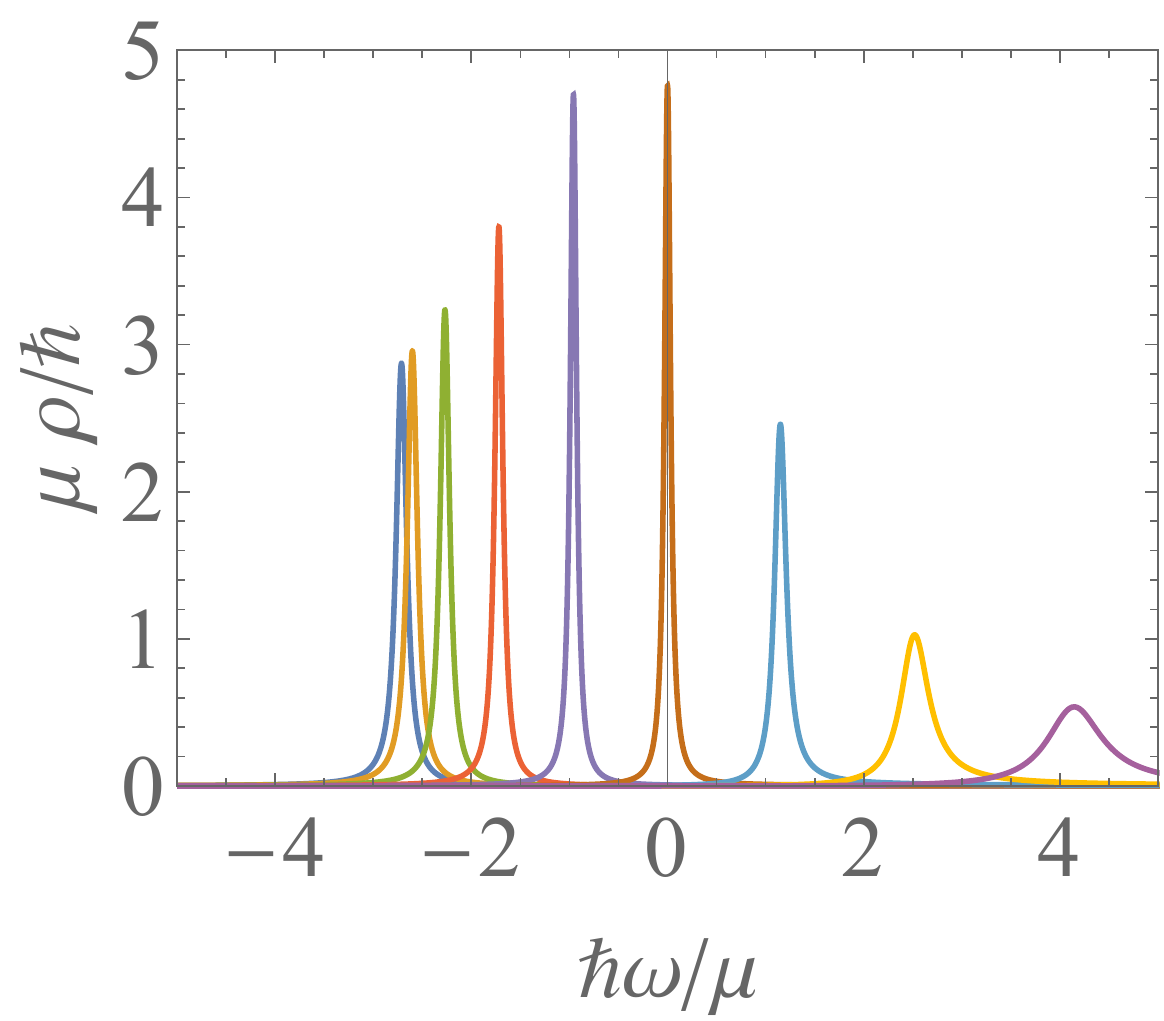}}
\caption{(color online) (a) Starting with a relativistic spectrum with particle and antiparticle peaks, we obtain nonrelativistic spectra by inspecting the nonrelativistic regime at small $\mu$, where the antiparticle peak decouples. (b) The spectral function at $\beta\mu=2$ and $k=6k_F/5$, for $k_{\rm B}T/M_0 c^2$ equal to $10^{-4}$ (blue), $2\cdot 10^{-4}$ (yellow), $3 \cdot 10^{-4}$ (green), $5\cdot 10^{-4}$ (red) and $10^{-3}$ (purple). (c) The universal spectral functions at $\beta\mu=2$ for several values of $k$. For the curve corresponding to the left peak, $k=0$, and $k$ increases by $k_F/5$ for successive curves. In both (b) and (c) we used $\{q,M,\lambda,g\}=\{1,49/100,-3/4,1/3\}$ \cite{Note4}, for which $\hbar k_F \simeq 2.1\sqrt{\mu M_0}$ and $Z\simeq 0.3$.}
\label{fig:nrlimit}
\end{figure}
The above (semi)holographic model yields relativistic spectral functions that obey the frequency sum rule $\int d\omega \rho(\omega,\vec{k})=4$ \cite{Arpes} and thus contain both particle and antiparticle peaks, separated by a gap proportional to $M_0 c^2$ \cite{Plantz2}. The introduction of this mass scale allows us to inspect the nonrelativistic limit by considering temperatures and chemical potentials that are small compared to this scale. For this, however, we first need a suitable definition of the nonrelativistic chemical potential $\mu$, which differs from $\mu_0$ defined above as in the limit $c\rightarrow \infty$ we want to measure the chemical potential with respect to the bottom of the particle band as illustrated in Fig. \ref{fig:nrlim}. Hence, when $\mu=T=0$ we expect a delta peak at $(\omega,\vec{k})=(0,\vec{0})$. Defining $\mu_0^*$ as the value of $\mu_0$ at which this occurs, the nonrelativistic chemical potential $\mu$ is then proportional to $\mu_0-\mu_0^*$.

Moreover, for a genuine nonrelativistic spectrum, we should observe that in the regime where $\hbar\omega$ and $\hbar ck$ are small compared to $M_0 c^2$, the spectral functions no longer depend on the energy scale $M_0 c^2$. Another way of saying this is that the spectra should only depend on the ratio $\beta\mu$ rather than on $k_{\rm B} T/M_0 c^2$ and $\mu/M_0 c^2$ separately. An obvious strategy to find such spectra is therefore to analyze spectral functions for several small values of $T$ and $\mu$, keeping the ratio $\beta\mu$ fixed. Our numerical data shown in Fig. \ref{fig:collapse} reveals that we can indeed find a data collapse in this limit, provided that we use the nonrelativistic chemical potential $\mu=Z(\mu_0-\mu_0^*)$, with the wavefunction renormalization factor $Z$ defined by $1/Z=-2\partial_\omega\text{Re}\left[\text{Tr\,}G_R(\omega,\vec{0})\right]^{-1}\big|_{\omega=0}$. In Fig. \ref{fig:spectralfunctions} we show the spectral functions obtained for $\beta\mu=2$ for several values of $k$. The locations $\omega(k)$ of the peaks in these spectra indeed conform to a nonrelativistic dispersion $\omega(k)=\hbar(k^2-k_F^2)/2M_{\text{eff}}$ with $k_F$ the Fermi momentum and $M_{\text{eff}} \simeq 0.86M_0$ for the model parameters in the figure that we have chosen such that the spectral functions resemble those of the unitary Fermi gas.

Finally, it is very important to realize that in principle the antiparticle part of the spectrum is still present in our numerics due to the fact that we can make the scale $M_0 c^2$ very large, but not truly infinite. However, this part must not be included in the nonrelativistic spectral function that only describes the particles. Naturally, this part of the spectrum also does not collapse. In practice this means that we should cut off the spectrum at some point inside the mass gap. Our results are not very sensitive to this cutoff, provided the scale $M_0 c^2$ is taken large enough. By construction, the final spectral functions then also satisfy the desired frequency sum rule $\int\dd\omega \rho(\omega,\vec{k})=2$ for spin-1/2 particles.

\textit{Unitary fermions.}---
Unitary fermions constitute, similar to the findings above, a system described at zero temperature by a set of universal constants and whose dimensionless thermodynamic functions depend solely on $\beta\mu$. An example of the former is the constant $\beta_{\rm SF}$ defined by $\mu=(1+\beta_{\rm SF})\epsilon_F$, with $\epsilon_F$ the Fermi energy. Experiments as well as theoretical models have determined that at zero temperature, so in the superfluid phase, $\beta_{\rm SF} \simeq -0.6$ \cite{DeSilva,Ku,Diener,Astrakharchik,Chin,Salomon}. The same quantity in the normal phase should in principle be slightly less negative, but is not accurately known at present. Therefore we have for simplicity taken our model parameters such that also $\beta_{\rm N} \simeq -0.6$. To see this from our spectra we can use that $\epsilon_F=\hbar^2 k_F^2/2m_{\text{id}}$ with $m_{\text{id}}$ the mass of the ideal Fermi gas. The Fermi momentum $\hbar k_F \simeq 2.1\sqrt{\mu M_0}$ follows directly from the dispersion in our spectral functions at low temperatures and the value of the mass $m_{\text{id}} \simeq 0.94 M_0$ we obtain from the dispersion of the critical system near $\mu=T=0$, since our spectral functions indeed contain a very sharp peak in this case.
\begin{figure*}[!t]
\centering
\subfloat[\label{fig:md}]
	{\includegraphics[trim = 0cm 0cm 0cm 0cm, clip=true, scale=.47]{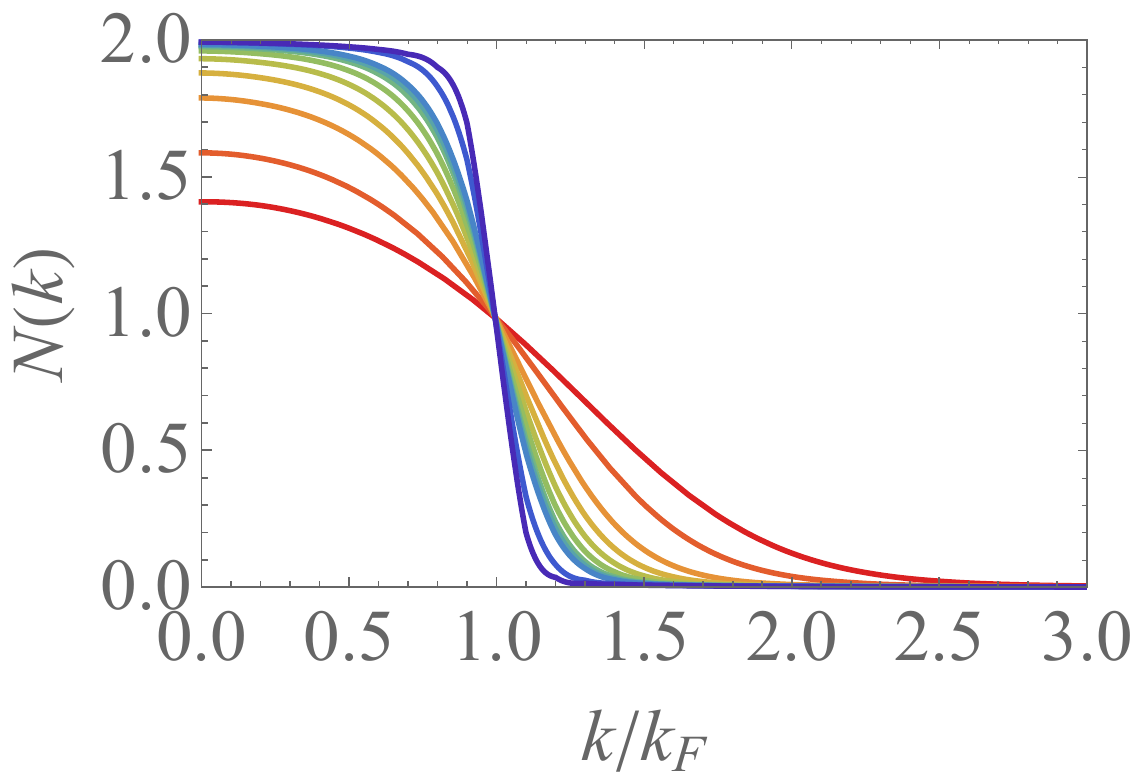}}
\hspace{0.8cm}
\subfloat[\label{fig:md4}]
	{\includegraphics[trim = 0cm 0cm 0cm 0cm, clip=true, scale=.47]{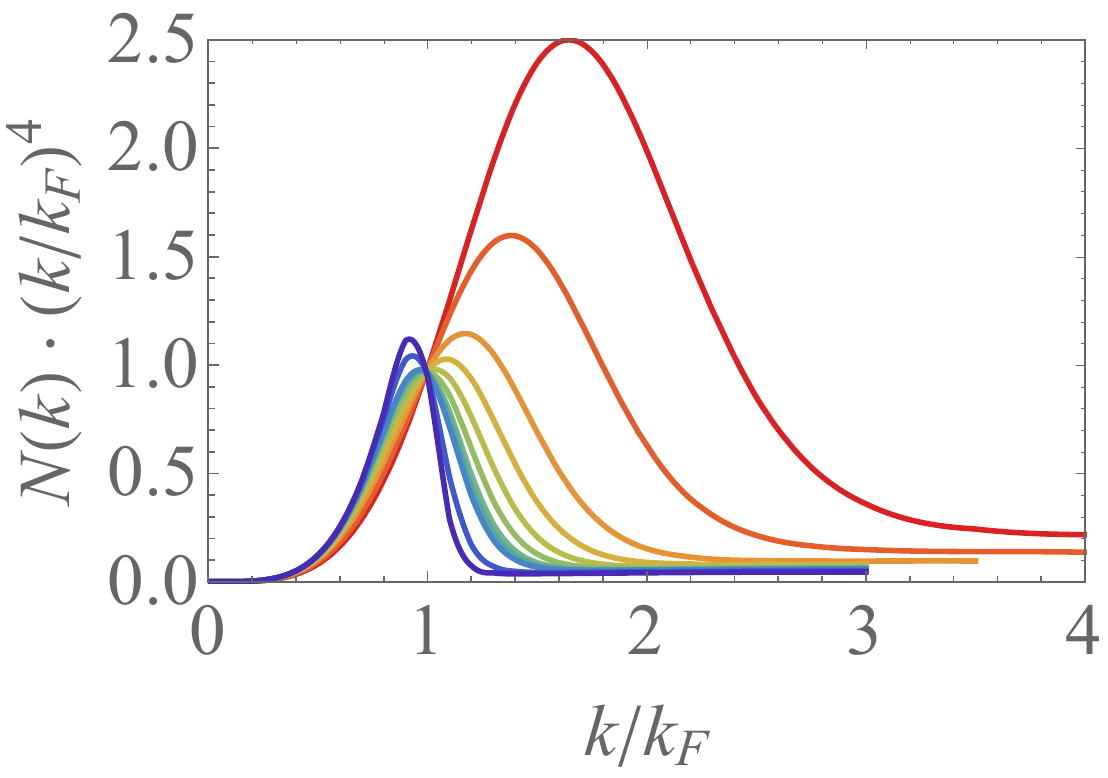}}
\hspace{0.8cm}
\subfloat[\label{fig:contact}]
	{\includegraphics[trim = 0cm 0cm 0cm 0cm, clip=true, scale=.4]{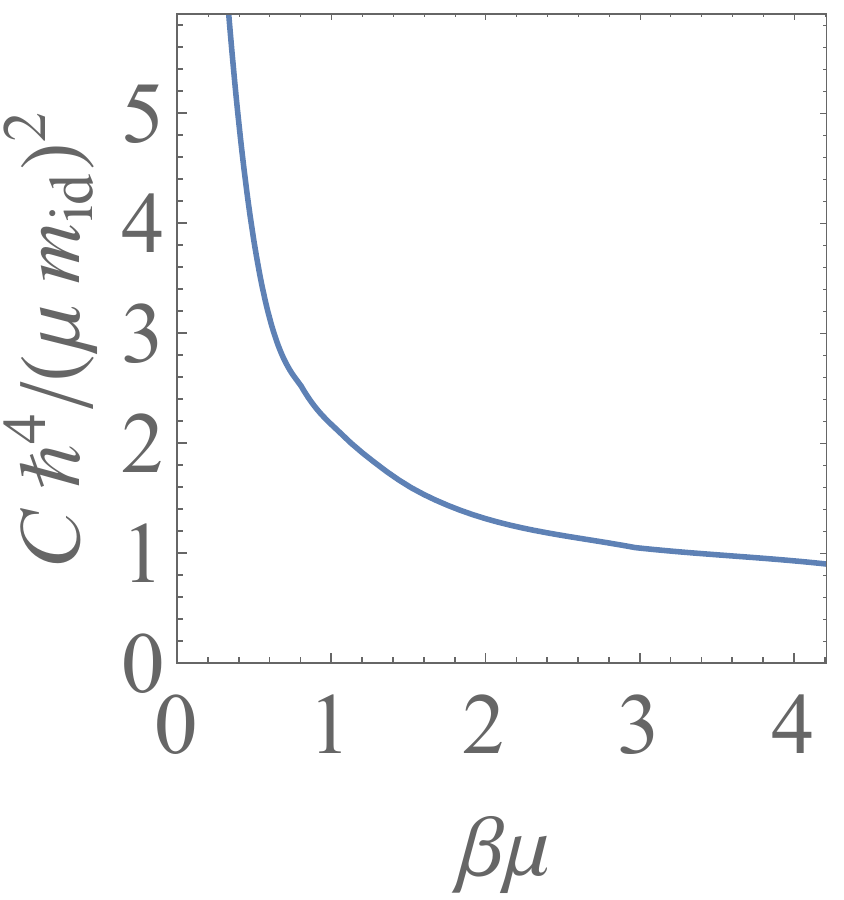}}
\caption{(color online) (a) The momentum distributions found from the spectral functions integrated with the minimum cut-off frequency $\omega=-10\mu/\hbar$ and including a wavefunction renormalization factor $1/Z$. In order of increasing steepness at $k_F$, the values of $\beta\mu$ are $0.34$, $0.51$, $0.79$, $1.03$, $1.27$, $1.51$, $1.75$, $1.87$, $1.99$, $2.97$ and $4.21$.  (b) The momentum distributions behave as $1/k^{4}$ for large $k$, although there are some deviations if $k$ is too large. The coloring is the same as in (a). (c) The contact parameter scaled with $\mu^2m_{\rm id}^2/\hbar^4$, as determined from the large momentum tails in (b).}
\label{fig:1}
\end{figure*}

From our spectral functions we can next calculate the momentum distributions $N({\bf k})=\int\dd\omega\rho(\omega,{\bf k}) n_F(\omega)$ with $n_F$ the Fermi-Dirac distribution. These can ultimately be used to determine the equation of state that was already shown previously in Fig.~1. Performing the calculation, whose outcome is shown in Fig.~\ref{fig:md}, we need to realize that the above-mentioned wavefunction renormalization requires us to add an additional factor of $1/Z$ to the momentum distribution. In this manner the integral of $N({\bf k})$ over momentum space approaches the ideal result at low temperatures, as expected from Luttinger's theorem. To understand also physically why this factor is necessary, we note that the spectral functions we obtain generically consist of the nonrelativistic low-frequency peak with a spectral weight of $2Z$ and a long tail at higher frequencies that contains most of the remaining spectral weight. Such tails are a recurring feature in holographic spectra and are related to the non-analytic behavior $(-\omega^{2}+c^2k^2)^M$ of the self-energy due to the asymptotically AdS gravitational background. This tail persists even in the low-temperature limit, where the momentum distribution only captures the states at small negative frequencies. We can therefore interpret our spectra as containing a `coherent' part of weight $2Z$, which is shown in Fig.~3 and describes the nonrelativistic unitary fermions, and an `incoherent' part of weight $2(1-Z)$.

Comparing the momentum distributions to the results of Ref.~\cite{Werner}, we see that many features of the unitary Fermi gas are reproduced by our nonrelativistic fermions. For instance, we have checked that the slope of $N({\bf k})$ at the Fermi momentum does not diverge in the zero-temperature limit, which signals a non-Fermi-liquid behavior and is in agreement with Fig.~6 of Ref.~\cite{Werner}. Moreover, the characteristic asymptotic behavior of the momentum distributions as $C/k^4$ for large $k$ in terms the contact parameter $C$, is also seen in our data in Fig.~\ref{fig:md4}, although we observe some relativistic corrections for even larger $k$. From this figure we can also read off the contact parameter, which results in Fig.~\ref{fig:contact}. We remind that to calculate the momentum distributions, we must introduce a frequency cutoff inside the gap from which we start integrating the spectral functions. The dependence on this cutoff is negligible for the momentum distributions in Fig.~\ref{fig:md} themselves, but the cutoff does have some influence on the tails in Fig.~\ref{fig:md4}, as small deviations get amplified by the factor $k^4$. Given these uncertainties and the fact that we can still fine-tune several model parameters, we find the agreement with Fig.~2b of Ref.~\cite{Werner} rather encouraging. Note that our momentum distributions contain also an interesting crossing point, which appears to be a universal feature of strongly interacting systems \cite{Vollhardt}.

Finally, we calculate the total density by integrating the momentum distributions over momentum space. The result, normalized by the ideal Fermi gas density, is shown in Fig. \ref{fig:density}. For low temperatures our result asymptotically approaches $1/(1+\beta_{\rm N})^{3/2}$ with $\beta_{\rm N} \simeq -0.6$, as expected. This limit is not clearly visible in the data of Refs.~\cite{Ku,Rammelmuller}, since there at low temperatures the unitary gas becomes superfluid, a feature that we have not included yet but can also be achieved holographically.

\textit{Discussion and outlook.}---
Up to now, we have used holography as a bottom-up approach in which the various model parameters can be tuned to fit to experiments. However, the holographic dictionary also provides insight into the physical significance of these parameters. For instance, the conformal dimension of the operator $\mathcal{O}$ is equal to $2+M$, and $q$ and $\lambda$ determine the strength of the CFT three-point functions
$\langle \mathcal{O}^\dagger \mathcal{O} (\Psi^\dagger\Psi) \rangle$ and
$\langle \mathcal{O}^\dagger \mathcal{O} (\Psi^\dagger\gamma^0\Psi) \rangle$, respectively. In the context of the unitary Fermi gas a natural choice for the operator $\mathcal{O}$ is $\Psi^\dagger$ multiplied with the annihilation operator of a Feshbach molecule. If this identification is correct then $g$ would correspond to the atom-molecule coupling of the Feshbach resonance. Exploring these possible microscopic connections in detail is beyond the scope of the present paper and is left for future work. Continuing in the spirit of bottom-up holography, however, our approach allows for many extensions by adding more ingredients to the gravitational dual theory, such as the inclusion of the backreaction on the bulk geometry by a complex scalar field that is dual to the superfluid order parameter and the introduction of a spin and/or mass imbalance.

\acknowledgments

This work was first presented at the Aachen workshop ``Quantum Many-Body Methods in Condensed Matter Systems''. It is a pleasure to thank the participants of this workshop for helpful discussions and feedback. Moreover, we thank Stefan Vandoren and Umut G\"ursoy for very stimulating discussions. This work was supported by the Stichting voor Fundamenteel Onderzoek der Materie (FOM) and is part of the D-ITP consortium, a program of the Netherlands Organisation for Scientific Research (NWO) that is funded by the Dutch Ministry of Education, Culture and Science (OCW).

\bibliographystyle{apsrev4-1}
\bibliography{Bibliography}
\onecolumngrid


\section*{Supplementary material}

# Supplemental Material for "Nonrelativistic fermions with holographic interactions and the unitary Fermi gas"

N.W.M. Plantz and H.T.C. Stoof
*Institute for Theoretical Physics and Center for Extreme Matter and Emergent Phenomena,
Utrecht University, Princetonplein 5, 3584 CC Utrecht, The Netherlands*[*]

(Dated: October 23, 2018)

## CONTENTS

## I. ACTIONS AND EQUATIONS OF MOTION FOR THE BULK THEORY

In this section we present more details on the gravitational background that is used to obtain the results in the main text. In particular, we present the equations of motion that need to be solved in order to obtain this background. Moreover, we present the equations of motion for the probe Dirac spinors propagating on this background, which ultimately lead to the self-energy of our spectral functions.

### A. Gravitational background

The bulk theory contains a gauge field $A = A_t \mathrm{d}t$ to account for the chemical potential in the CFT and a scalar field $\phi$ to account for the mass deformation in the CFT. The scalar field is tachyonic with mass $m_\phi^2 = -3$, such that the corresponding deformation of the CFT has the dimension of a fermionic mass deformation.

The gravitational background follows from the backreaction of these fields on the geometry described by the metric $g_{MN}$, which follows from the action

$$S_{\mathrm{background}} = \int \mathrm{d}^5 x \sqrt{-g} \left( R + 12 - \frac{1}{4}F^2 - \frac{1}{2}\left((\partial\phi)^2 + m_\phi^2 \phi^2\right) \right). \quad (1)$$

Here, $g$ is the determinant of the metric, $R$ is the Ricci scalar, $F = \mathrm{d}A$ and $(\partial\phi)^2 = \partial_M \phi \partial^M \phi$. Moreover, we note that the first two terms in the Lagrangian density represent the standard Einstein-Hilbert Lagrangian $R - 2\Lambda$, since in our units the cosmological constant is given by $\Lambda = -6$ as explained in Appendix A.

For the metric $\mathrm{d}s^2 = g_{MN} \mathrm{d}x^M \mathrm{d}x^N$ we use the following *Ansatz*:

$$\mathrm{d}s^2 = -f(r)e^{-\chi(r)} \mathrm{d}t^2 + \frac{\mathrm{d}r^2}{f(r)} + r^2 \mathrm{d}\mathbf{x}^2, \quad (2)$$

---

[*] n.w.m.plantz@uu.nl; h.t.c.stoof@uu.nl

where the metric components as well as $A_t$ and $\phi$ only depend on the radial coordinate $r$ due to planar symmetry. The equations of motion following from Eq. (1) can then be written as

$$\phi'' + \left(\frac{f'}{f} + \frac{3}{r} - \frac{\chi'}{2}\right)\phi' + \frac{3}{f}\phi = 0, \tag{3}$$

$$A_t'' + \left(\frac{3}{r} + \frac{\chi'}{2}\right)A_t' = 0, \tag{4}$$

$$\chi' + \frac{r}{3}\phi'^2 = 0, \tag{5}$$

$$f' + \left(\frac{2}{r} - \frac{\chi'}{2}\right)f + \frac{r}{6}e^\chi A_t'^2 - \frac{r}{2}\phi^2 - 4r = 0, \tag{6}$$

where a prime denotes differentiation with respect to $r$. A gravitational background then follows from solving this system with the initial conditions $f(r_+) = 0$, $A_t(r_+) = 0$, $\chi(r_+) = 0$ and two free initial conditions $\phi(r_+)$ and $A_t'(r_+)$. After solving the system, the solution is rescaled such that in the end $\chi(\infty) = 0$ as required for asymptotically anti-de Sitter spacetimes. It can be shown that $\phi'(r_+)$ is not independent of the other initial conditions.

After numerically solving the above system of equations, we extract the parameters of the CFT. The temperature follows from the metric tensor via

$$T = \frac{f'(r_+)e^{-\chi(r_+)/2}}{4\pi}, \tag{7}$$

whereas the chemical potential per unit charge and the mass are given by the boundary values $\mu_0/q = A_t(\infty)$ and $M_0 = \lim_{r\to\infty} r\phi(r)/\alpha$ respectively. Here the proportionality constant $\alpha = \sqrt[4]{3/\pi^2}$ is discussed in Ref. [1], however, note that there $\alpha$ is defined as what is $1/\alpha$ here.

### B. Probe spinors

The self-energy of our spectral functions follow from the solution $\Xi$ of Eq. (3) in the main text. To derive this equation, we have two Dirac spinors $\psi^{(1)}$ and $\psi^{(2)}$ propagate on the bulk theory obtained from the equations of motion above. These spinors have masses $M_1 = M$ and $M_2 = -M$ respectively and are coupled to the gauge field $A_M$ with a charge $q$. The associated action is given by

$$S_{\text{Dirac}} = ig_f \int d^5x\sqrt{-g}\left(\bar\psi^{(1)}\left(\slashed{D} - M\right)\psi^{(1)} + \bar\psi^{(2)}\left(\slashed{D} + M\right)\psi^{(2)}\right)$$
$$+ ig_Y \int d^5x\sqrt{-g}\phi\left(\bar\psi^{(1)}\psi^{(2)} + \bar\psi^{(2)}\psi^{(1)}\right) + ig_f \int d^4x\sqrt{-h}\left(\bar\psi_R^{(1)}\psi_L^{(1)} - \bar\psi_L^{(2)}\psi_R^{(2)}\right). \tag{8}$$

where $\bar\psi = \psi^\dagger \Gamma^{\underline{0}}$, $\slashed{D} = \Gamma^M(\nabla_M - iqA_M)$, $g_f$ and $g_Y$ are coupling constants, $h$ is the determinant of the induced metric on the boundary and $\psi_{R,L}^{(i)} = (1 + \Gamma^{\underline{r}})\psi^{(i)}/2$. The spinor covariant derivative $\nabla$ and the Dirac matrices in $(4+1)$-dimensional flat ($\Gamma^{\underline{M}}$) and curved ($\Gamma^M$) spacetime are defined in Appendix A. The action consists of a standard Dirac action for the spinors $\psi^{(i)}$, a Yukawa term which is necessary to couple the chiral components of the spinor on the boundary and a boundary action to be consistent with the Dirichlet boundary conditions $\delta\psi_R^{(1)} = 0$ and $\delta\psi_L^{(2)} = 0$. Defining $\lambda = g_Y/g_f$, the equations of motion from the spinor are then

$$\left(\slashed{D} - M\right)\psi^{(1)} = -\lambda\phi\psi^{(2)}, \tag{9}$$

$$\left(\slashed{D} + M\right)\psi^{(2)} = -\lambda\phi\psi^{(1)}. \tag{10}$$

Next, we define the Dirac spinors $\Psi = \psi_R^{(1)} + \psi_L^{(2)}$ and $\eta = \psi_L^{(1)} - \psi_R^{(2)}$, in terms of which the on-shell action is

$$S^{\text{on shell}} = ig_f \int d^4x\sqrt{-h}\bar\Psi\eta. \tag{11}$$

The matrix $\Xi$ is now defined in momentum space by

$$\eta(r,k) = -i\Xi(r,k)\Psi(r,k), \tag{12}$$

so that $\Xi$ is related to the Green's function for the fermionic boundary operator sourced by the Dirac spinor $\Psi$ on the boundary. Eq. (3) in the main text then follows from the above definition when imposing the Dirac equations for $\Psi$ and $\eta$, which follow from rewriting the Dirac equations for $\psi^{(1)}$ and $\psi^{(2)}$.

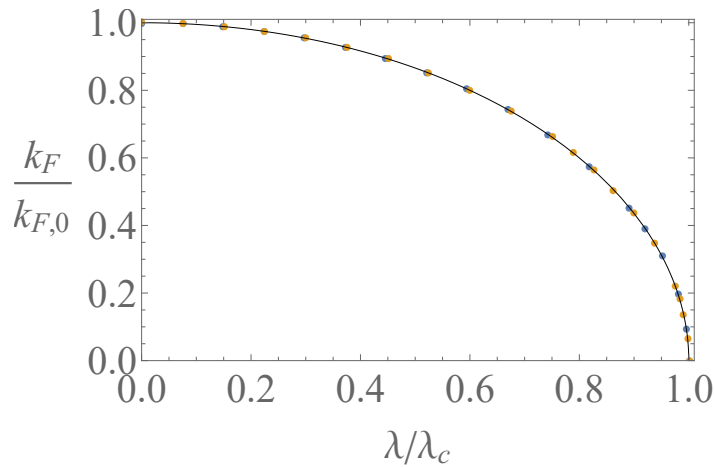

FIG. 1. The Fermi momentum at $g = \infty$ as a function of $\lambda$ depends on $\mu_0$ only through $k_{F,0}$ and $\lambda_c$. Here, we used $q = 1$, $M = 49/100$ and $k_B T = 10^{-4} M_0 c^2$, which is small enough to determine $k_F$. The values of $\mu_0/M_0 c^2$ are $1/2$ (blue dots) and $1$ (yellow dots). The black curve shows the graph of $\sqrt{1-x^2}$, with $x = \lambda/\lambda_c$.

## II. ON THE CHOICE OF PARAMETERS USED TO OBTAIN NONRELATIVISTIC SPECTRA

In general, a spectral function $\rho(\omega, \mathbf{k})$ depends on the gravitational-background parameters $k_B T/M_0 c^2$ and $\mu_0/M_0 c^2$ and the model parameters $q$, $M$ and $\lambda$ and $g$. Not every set of the parameters $\{q, M, \lambda, g\}$ is suitable to obtain universal nonrelativistic spectra with holographic interactions. Firstly, to satisfy the frequency sum rule, we must have that $-1/2 < M < 1/2$ [2]. Moreover, we can restrict to positive $g$ since the spectral functions only depend on $g^2$. In this section we discuss some more restrictions on this set, which we have taken into account for the values of the paramaters used in the main text. In particular, fixing $q$ and $M$, we find a restriction on $\lambda$.

To derive such restrictions, we should realize that the self-energy contains a gap itself. If the peaks in the nonrelativistic spectral functions are situated inside this gap, they will not be broadened and the resulting spectrum will resemble a noninteracting one, containing delta peaks at each value of $k$. Since the gap in the self-energy $\Xi$ is proportional to $|\lambda| M_0 c^2$, we expect this to occur for large values of $|\lambda|$. In the analysis below we indeed find an upper bound for $|\lambda|$.

For $k = 0$, the peak in the nonrelativistic spectrum is not situated inside the gap of the self-energy if we restrict to chemical potentials $\mu_0$ that are greater than the critical chemical potential $\mu_0^*$ in the limit $g \to \infty$. Since for nonrelativistic spectra $\mu_0 \simeq \mu_0^*$, we can write this criterion as

$$\mu_0^*(q, M, \lambda, g) > \mu_0^*(q, M, \lambda, \infty). \tag{13}$$

This condition should also be sufficient for nonrelativistic spectral functions at nonzero $k$, provided that the difference between $\mu_0^*(q, M, \lambda, g)$ and $\mu_0^*(q, M, \lambda, \infty)$ is not nonrelativistically small. For parameters satisfying this condition, we indeed find spectra containing peaks with a nontrivial width, such as the ones in the main text.

To see what the above condition implies for the allowed sets of model parameters, we study the behavior of the critical chemical potential $\mu_0^*$ as a function of $\lambda$ and $g$ for fixed $q$ and $M$. Noting that at the critical chemical potential we have that $k_F = 0$, we can use that the Fermi momentum $k_F(\lambda, g, \mu_0)$ at $g = \infty$ depends on $\lambda$ as

$$k_F(\lambda, \infty, \mu_0) = k_{F,0} \sqrt{1 - \frac{\lambda^2}{\lambda_c^2}}, \tag{14}$$

where $k_{F,0} = k_F(0, \infty, \mu_0)$ and $\lambda_c$ is defined as the positive value of $\lambda$ at which the Fermi momentum at $g = \infty$ is zero. This dependence is found numerically and is shown in Fig. 1. All dependence on $\mu_0$ is contained in $k_F(0, \infty, \mu_0)$ and $\lambda_c(\mu_0)$. Putting Eq. (14) to zero yields that the critical chemical potential at $g = \infty$ is given by the solution of $|\lambda| = \lambda_c(\mu_0)$. It was furthermore found in Ref. [1] that except for small $\mu_0$, both $k_{F,0}$ and $\lambda_c$ are linear in $\mu_0 > 0$, so

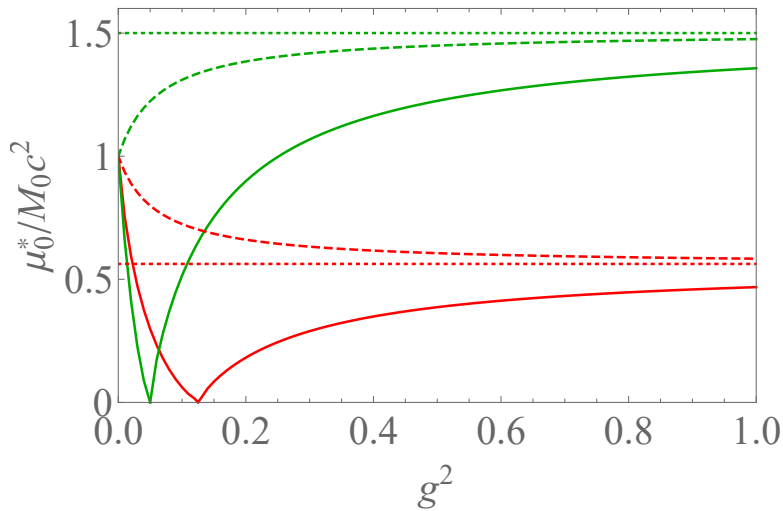

FIG. 2. The critical chemical potential as a function the coupling $g$. Here $|\lambda| = 3/4$ for the red curves and $|\lambda| = 2$ for the green curves. The solid curves correspond to positive $\lambda$ and the dashed curves correspond to negative $\lambda$. The dotted lines denote the asymptotic values of the curves, which are equal to $|\lambda|/B$. Here we used $q = 1$ and $M = 49/100$, for which $B \simeq 1.33$.

that in this regime we can write $\lambda_c \simeq B\mu_0/M_0 c^2$ with $B$ a positive coefficient which depends on $q$ and $M$. It follows that at $g = \infty$ we get

$$\mu_0^*(\lambda, \infty) = \frac{|\lambda|}{B} M_0 c^2 \tag{15}$$

so that the criterion in Eq. (13) can be written as $|\lambda|M_0 c^2 < B\mu_0^*(\lambda, g)$.

We proceed by studying the dependence of the critical chemical potential $\mu_0^*$ on $\lambda$ and $g$, of which the result is shown in Fig. 2. Clearly, for $g = 0$ we have that $\mu_0^* = M_0 c^2$. For $\lambda < 0$, we then find that $\mu_0^*/M_0 c^2$ is a monotonic function starting at 1 and asymptotically approaching $|\lambda|/B$. In contrast, when $\lambda > 0$ we observe that $\mu_0^*/M_0 c^2$ monotonically decreases to 0 for some value of $g$, after which it monotonically increases to $\lambda/B$. These findings indicate that we can only obey the criterion in Eq. (13) if $\mu_0^*(\lambda, \infty) < M_0 c^2$, i.e.,

$$|\lambda| < B, \tag{16}$$

which for fixed values of $q$ and $M$ gives an upper bound for $|\lambda|$. From Fig. 2 we furthermore observe that there is no restriction on $g$ for negative $\lambda$, whereas for positive $\lambda$ an upper bound for $g^2$ is required to satisfy Eq. (16).

For the parameters $q = 1$ and $M = 49/100$, which are used in the main text, we have that $B \simeq 1.33$, so that our choice $\lambda = -3/4$ satisfies the criterion above in Eq. (16). We remark that we should not choose $|\lambda|$ too small, since then an additional peak near the chemical potential at $\omega = 0$ appears in our spectra. This leads to an avoided crossing between this peak and the particle band, which makes it hard to define the critical chemical potential. Having opted for a negative $\lambda$, the above discussion imposes no further restriction on $g$. However, to obtain nonrelativistic spectra, we should not take $g$ too large. To see this, we note that we can think of $g$ as a measure for the region in momentum space where the holographic interactions dominate the free kinetic part of the Green's function. As a consequence, for large $g$ we find that the tails at higher frequencies in the spectral functions that were mentioned in the main text persist until deep in the relativistic regime. Moreover, the spectral weight in the gap is then no longer negligible. This is not the case for the value $g = 1/3$ used in the main text, which we have taken to reproduce the value $\beta_N \simeq -0.6$. Finally, the chosen value for $M$ in the main text is also related to the above-mentioned tails. We find that choosing $M$ close to its supremum $1/2$ avoids long tails extending to relativistic $\omega$. In particular, the spectral functions decay faster than $1/\omega$ for large $\omega$, i.e., for $\omega \gg \mu$ but still within the nonrelativistic regime, as is necessary to obey the sum rule.

**Appendix A: Conventions on units and Dirac theory**

The action for the gravitational background in Eq. (1) in SI units reads

$$S = \int d^5x \sqrt{-g} \left( \frac{c^3}{16\pi G_5}(R - 2\Lambda) - \frac{1}{4\mu_5 c} F^2 - \left( (\partial\phi)^2 + \frac{m_\phi^2 c^2}{\hbar^2}\phi^2 \right) \right). \tag{A1}$$

Here $G_5$ and $\mu_5$ are Newton's constant and the vacuum permeability respectively, defined in 4+1 spacetime dimensions. Using the cosmological constant $\Lambda$ we can define the anti-de Sitter radius as $L^2 = 6/(-\Lambda)$, noting that the cosmological constant is always negative in the asymptotically anti-de Sitter spacetimes that we are dealing with. The dimensionless gauge field and scalar field are then defined as

$$\tilde{A}_{\underline{t}} = \sqrt{\frac{16\pi G_5}{\mu_5 c^6}} A_t, \tag{A2}$$

$$\tilde{\phi} = \sqrt{\frac{16\pi G_5}{c^3}} \phi. \tag{A3}$$

The tildes, which we omit in the main text, denote dimensionless quantities. The metric signature is mostly plus and its components are defined by $ds^2 = g_{MN} dx^M dx^N$, with $x^M = \{ct, r, \mathbf{x}\}$ where capital Roman letters refer to $(4+1)$-dimensional spacetime, as opposed to Greek letters for which $x^\mu = \{ct, \mathbf{x}\}$. With this definition of the metric, the components $g_{MN}$ are already dimensionless. All dimensionless units in main text are obtained by scaling all length scales by $L$, i.e., putting $\Lambda = -6$. As a consequence, when putting $\hbar = 1$ and $c = 1$, all energy (or mass) scales are expressed in units of $\hbar c/L$ (or $\hbar/cL$). This is also true for the temperature $T$, setting Boltzmann's constant $k_B = 1$. Finally, the Dirac fields in Eq. (8) are in units of $\sqrt{\hbar}/L$ and the dimensionless charge of the probe field is given by

$$\tilde{q} = \sqrt{\frac{\mu_5 c^6}{16\pi G_5}} \frac{L}{\hbar c} q. \tag{A4}$$

In the main text we use the dimensionless units defined here for bulk parameters such as $M$ and $q$. For quantities defined in the CFT we use SI units, which means we restore $c$, $\hbar$ and $k_B$.

The Dirac matrices in flat $(3+1)$-dimensional spacetime are given by

$$\gamma^{\underline{\mu}} = \begin{pmatrix} 0 & \bar{\sigma}^\mu \\ \sigma^\mu & 0 \end{pmatrix} \tag{A5}$$

where $\sigma = (\mathbb{1}_2, \sigma^i)$ and $\bar{\sigma} = (-\mathbb{1}_2, \sigma^i)$ with $\sigma^i$ the Pauli matrices and $\mathbb{1}_2$ the $2 \times 2$ identity matrix. Moreover, we use underlined indices for tensors and Dirac matrices defined in flat spacetime, so that $g^{\underline{MN}} = \eta^{\underline{MN}} = \text{diag}(-1, 1, 1, 1, 1)$. The gamma matrices $\Gamma^{\underline{M}}$ in $(4+1)$-dimensional flat spacetime are given by $\Gamma^{\underline{\mu}} = \gamma^{\underline{\mu}}$ for $\mu \neq r$ and

$$\Gamma^{\underline{r}} = \gamma^{\underline{5}} \equiv i\gamma^{\underline{0}}\gamma^{\underline{1}}\gamma^{\underline{2}}\gamma^{\underline{3}} = \begin{pmatrix} \mathbb{1}_2 & 0 \\ 0 & -\mathbb{1}_2 \end{pmatrix}. \tag{A6}$$

The vielbeins $e^M_{\underline{N}}$ that appear in the Dirac action in Eq. (8) in curved spacetime are defined by

$$g_{MN} = e^{\underline{P}}_M e^{\underline{Q}}_N \eta_{\underline{PQ}} \tag{A7}$$

where the inverse vielbeins satisfy $e^M_{\underline{P}} e^{\underline{P}}_N = \delta^M_N$ and $e^{\underline{P}}_M e^N_{\underline{P}} = \delta^{\underline{N}}_{\underline{M}}$. For the metric in Eq. (2) this gives

$$e^0_{\underline{0}} = \sqrt{\frac{e^{\chi(r)}}{f(r)}}, \tag{A8}$$

$$e^r_{\underline{r}} = \sqrt{f(r)}, \tag{A9}$$

$$e^i_{\underline{i}} = \frac{1}{r}. \tag{A10}$$

In the main text, we have omitted the underlines and only use the Dirac matrices in flat spacetime. Moreover, all vielbeins in the main text are such that their lower index corresponds to the flat one.

The spinor covariant derivative $\nabla_M$, which also appears in the Dirac action in Eq. (8), is defined as

$$\nabla_M \psi = \partial_M \psi + \Omega_M \psi \tag{A11}$$

with $\Omega_M$ given by

$$\Omega_M = \frac{1}{8} \omega_{M\underline{NP}} [\Gamma^{\underline{N}}, \Gamma^{\underline{P}}] \tag{A12}$$

and the spin connection $\omega^M_{N\underline{P}}$ given by

$$\omega^M_{N\underline{P}} = e^M_Q e^R_{\underline{P}} \Gamma^Q_{NR} - e^Q_{\underline{P}} \partial_N e^M_Q. \tag{A13}$$

Here $\Gamma^M_{NP}$ denotes the Christoffel connection. The spin connection does not appear in the equation for $\Xi$ in the main text, as we can remove it by rescaling the probe spinors by a function depending on $r$ only, see appendix A.2 in Ref. [1] for details.

---

12


\end{document}